\documentclass[10pt,twocolumn]{IEEEtran}
\usepackage{makeidx}
\usepackage{graphicx}
\usepackage{algorithm}
\usepackage{algorithmic}
\usepackage{authblk}
\usepackage{multirow,longtable}
\usepackage{amssymb}
\usepackage{color}
\usepackage{url}
\usepackage[utf8]{inputenc}
\usepackage{cancel}
\usepackage{tikz}
\newtheorem{definition}{Definition}
\usepackage{caption}
\usetikzlibrary{shapes.geometric, arrows}

\author{
    \IEEEauthorblockN{Bassam Alkindy\IEEEauthorrefmark{1}\IEEEauthorrefmark{3}, \and Jean-Fran\c{c}ois Couchot\IEEEauthorrefmark{1}\IEEEauthorrefmark{3}, \and Christophe Guyeux\IEEEauthorrefmark{1}\IEEEauthorrefmark{3} \and \\Arnaud Mouly\IEEEauthorrefmark{2}\IEEEauthorrefmark{3}, \and Michel Salomon\IEEEauthorrefmark{1}\IEEEauthorrefmark{3}, \and Jacques M. Bahi\IEEEauthorrefmark{1}\IEEEauthorrefmark{3}}\\
    \IEEEauthorblockA{\IEEEauthorrefmark{1}FEMTO-ST Institute, UMR 6174 CNRS, DISC Computer Science Department}\\
    \IEEEauthorblockA{\IEEEauthorrefmark{2}Chrono-Environnement Lab., UMR 6249 CNRS}\\
    \IEEEauthorblockA{\IEEEauthorrefmark{3}University of Franche-Comt\'{e}, France}
}

\title{Finding the Core-Genes of Chloroplasts}

\begin{document}

% make the title area
\maketitle

\begin{abstract}
Due to the recent evolution of sequencing techniques, the number of
available genomes is rising steadily, leading to the possibility to
make large scale genomic comparison between sets of close species.
An interesting question
to answer is:  
 what is  the common functionality genes of a collection
of species, or conversely, to determine what is specific to a given
species when compared to other ones belonging in the same genus, family, etc.
Investigating such  problem means to find both core and pan genomes
of a collection of species, \textit{i.e.}, genes in common to all the species
vs. the set of all genes in all species under consideration. However, obtaining
trustworthy core and pan genomes is not an easy task, leading to a large 
amount of computation, and requiring a rigorous methodology. Surprisingly,
as far as we know, this methodology in finding core and pan genomes has not really been
deeply investigated. This research work tries to fill this gap
by focusing only on chloroplastic genomes, whose reasonable sizes allow a deep study.
To achieve this goal, a collection of 99 chloroplasts are 
considered in this article. Two methodologies have been
investigated, respectively based on sequence similarities and 
genes names taken from annotation tools.
The obtained results will finally be evaluated in terms of biological relevance.
\end{abstract}

\begin{IEEEkeywords}
Chloroplasts,
Coding sequences,
Clustering,
Genes prediction,
Methodology,
Pan genome,
Core genome
\end{IEEEkeywords}

\section{Introduction} \label{sec:intro}

Identifying  core genes  may be of importance either to understand the shared functionality and specificity of a given set of species, or to construct their phylogeny using curated sequences. Therefore, in  this work we present methods to determine both core and pan genomes of a large set of DNA sequences. More precisely,  we focus on the following  questions by  using a  collection of 99~chloroplasts as an illustrative example: how can  we identify the  best core  genome (that is, an artificially designed set of functional  coding sequences as close as possible to the real biological one) and how to deduce scenarios regarding their genes loss. In other words, how to deduce scenarios regarding the gene increasing compared to the core genome?

Chloroplasts found in Eucaryotes have an endosymbiotic origin, which means that they come from the incorporation of a photosynthetic bacteria (Cyanobacteria) within an eucaryotic cell, which means they are the fundamental key elements in living organisms history, as they are organelles responsible for photosynthesis. This latter is the main way to produce organic matters from mineral ones using solar energy. Consequently photosynthetic organisms are at the basis of most ecosystem trophic chains. Indeed photosynthesis in Eucaryotes allow a great speciation in the lineage, leading to a great biodiversity. From an ecological point of view, photosynthetic organisms are at the origin of the presence of dioxygen in the atmosphere (allowing extant life) and are the main source of mid to long term carbon storage, which is fundamental regarding current climate changes. However, the chloroplasts evolutionary history is not totally well understood, at large scale, and their phylogeny requires to be further investigated.

A key idea in phylogenetic classification is that a given DNA mutation shared by at least two taxa has a larger probability to be inherited from a common ancestor than to have occurred independently~\cite{DBLP:journals/procedia/BahiGP12}. Thus shared changes in genomes allow to build relationships between species. In the case of chloroplasts, an important category of genomes changes is the loss of functional genes, either because they become ineffective or due to a transfer to the nucleus. Thereby a small number of gene losses  among species indicates that these species are close to  each other and belong to a similar lineage, while a  large loss  means distant lineages.

Phylogenies of photosynthetic plants are important to assess the origin of chloroplasts and the modes of gene loss among lineages. These phylogenies are usually done using a few chloroplastic genes, some of them being not conserved in all the taxa. This is why selecting core genes may be of interest for a new investigation of photosynthetic plants phylogeny. Such investigations have already been started in in [9], where core genome for photosynthetic productivity in \emph{Cyanobacteria} (\emph{Synechococcus} and \emph{Prochlorococcus}) has been regarded. 
Authors identified core photosystem II genes in cyanophages, which may increase viral fitness by supplementing the host production of some specific types of proteins.  
The study also proposed evidences of the presence of photosystem I genes in the genomes of viruses that affect cyanobacteria. However, the circumscription of the core chloroplast genomes for a given set of photosynthetic organisms needs bioinformatics investigations using sequence annotation and comparison tools, for which choices are available.

Our intention in this first research work regarding the methodology in core and pan genomes determination is to investigate the impact of these choices.  
A general presentation of the approaches detailed in this document is provided in the next section. Then we will  study in Section~\ref{sec:simil} the use of annotated genomes from NCBI website~\cite{Sayers01012011} with a coding sequences clustering method based on the Needleman-Wunsch similarity scores~\cite{Rice2000}. We will show that such an approach based on sequences similarity cannot lead to satisfactory results, biologically speaking. 
We will thus investigate name-based approaches in Section~\ref{sec:annot}, by using successively the gene names provided by NCBI and DOGMA~\cite{RDogma} annotations, where DOGMA is a recent annotation tool specific to chloroplasts. Finally, a discussion based on biological aspects regarding the evolutionary history of the considered genomes will finalize our investigations, leading to our methodology proposal for core and pan genomes discovery of chloroplasts. This research work ends by a conclusion section, in which our investigations will be summarized and intended future work will be planned.

\section{General presentation} \label{sec:general}
\begin{figure}[H]
\vspace{0.1cm} 
\centering
\includegraphics[width=.5\textwidth]{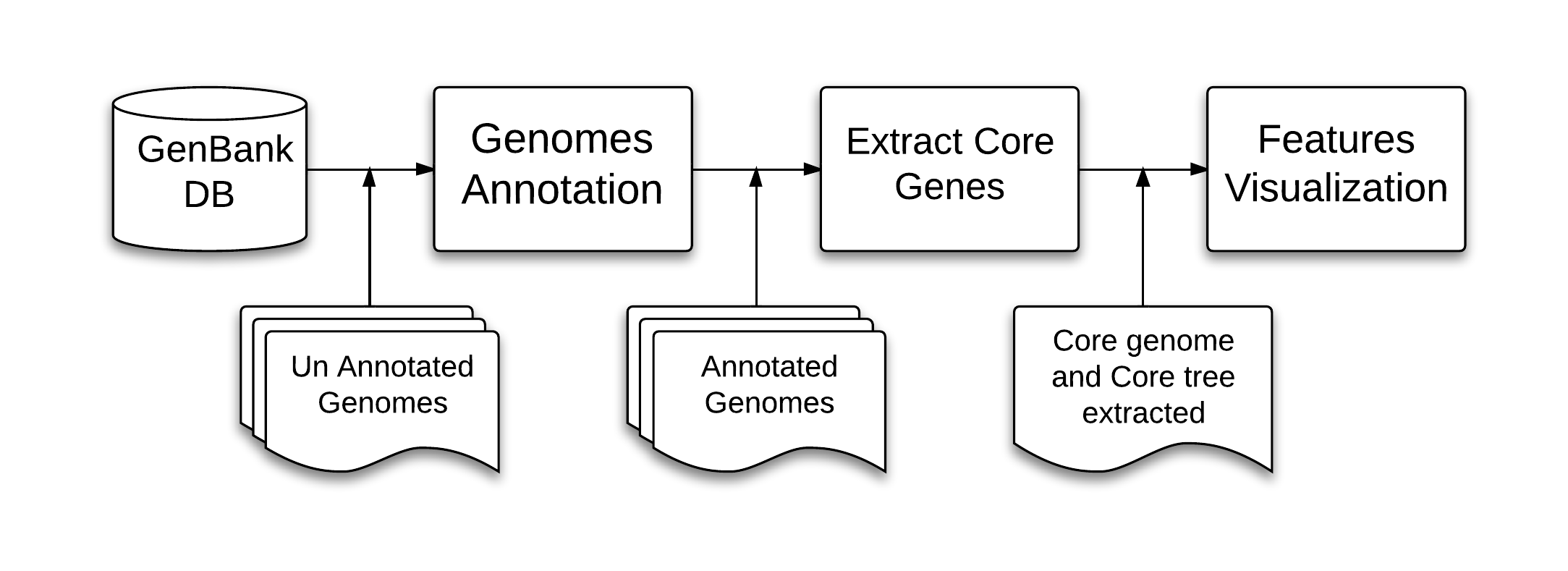}
\caption{A general overview of the annotation-based approach}\label{Fig1}
\end{figure}

Figure~\ref{Fig1} presents a general overview of the entire proposed pipeline
for core and pan genomes production and exploitation, which 
consists   of    three stages:   \textit{Genomes    annotation}, \textit{Core   extraction},
and \textit{Features Visualization}.  

As a starting  point, the pipeline uses a DNA sequences database like NCBI's GenBank~\cite{Sayers01012011}, the   European \textit{EMBL} database~\cite{apweiler1998swiss}, or the Japanese  \textit{DDBJ} one~\cite{sugawara2008ddbj}. 
It is possible to obtain annotated genomes (DNA coding sequences with gene names and locations) by interacting with these databases, either by directly downloading annotated genomes delivered by these websites, or by launching an annotation tool on complete downloaded  genomes. Obviously, this annotation stage must be of quality if we want to obtain acceptable core and pan genomes.
Various  cost-effective annotation  tools~\cite{Bakke2009} that produce genomical annotations at many detailed levels have been designed recently, some reputed ones being: 
DOGMA~\cite{RDogma},       cpBase~\cite{de2002comparative},
CpGAVAS~\cite{liu2012cpgavas},                   and
CEGMA~\cite{parra2007cegma}. Such tools usually use one out of the
three following methods  for finding  gene locations in large DNA sequences: 
\textit{alignment-based},  \textit{composition based},
or a  combination of both~\cite{parra2007cegma}.   The alignment-based
method  is used  when trying  to predict  a protein coding  sequence 
 by aligning a genomic DNA sequence with a
cDNA  sequence  coding  an already known homologous  protein~\cite{parra2007cegma}.
This approach is used for instance  in GeneWise~\cite{birney2004genewise}.  The
alternative   method,   the    composition-based   one   (also   known
as  \textit{ab initio})  is based  on  probabilistic  models of  genes
structure~\cite{parra2000geneid}. 

Using such annotated genomes, we will detail two general approaches for extracting the core genome, which is the second stage of the pipeline: the first one uses similarities computed on predicted coding sequences, while the second one uses all the information provided during the annotation stage. Indeed, such annotations can be used in various manners (based on gene names,  gene  sequences, protein  sequences, etc.) to extract the core and pan genomes.

The  final stage of our pipeline, only envoked in this article, is to take advantage of the information produced during the core and pan genomes search.
This features visualization stage encompasses phylogenetic tree construction using core genes, genes content evolution illustrated by core trees, functionality investigations, and so on.

For illustration purposes, we have considered 99~genomes of chloroplasts downloaded from GenBank database~\cite{Sayers01012011} as shown in Table~\ref{Tab2}. These genomes lie in  the eleven type  of chloroplast families. Furthermore, two kinds of annotations will be considered in this document, namely the ones provided by NCBI on the one hand, and the ones by DOGMA on the other hand.

  \begin{table}
    \tiny
    \caption[NCBI Genomes Families]{List of chloroplast genomes of photosynthetic Eucaryotes lineages from NCBI\label{Tab2}}
    \begin{minipage}{0.80\textwidth}
    \begin{minipage}{0.30\textwidth}
      \setlength{\tabcolsep}{4pt}
      \begin{tabular}{|p{0.1cm}|p{0.1cm}|p{1cm}|p{2cm}|}
      \hline
      {{F.}}&{{\#}} & {{Acc. No}} & {{Scientific Name}} \\ 
      \hline
	%Entering First line
      \parbox[t]{1mm}{\multirow{11}{*}{\rotatebox[origin=c]{90}{Brown Algae}}} &\multirow{11}{*}{11}  & NC\_001713.1 & \textit{Odontella sinensis} \\ 
      & & NC\_008588.1 & \textit{Phaeodactylum tricornutum} \\ 
      & & NC\_010772.1 & \textit{Heterosigma akashiwo} \\ 
      & & NC\_011600.1 & \textit{Vaucheria litorea} \\ 
      & & NC\_012903.1 & \textit{Aureoumbra lagunensis} \\ 
      &  & NC\_014808.1 & \textit{Thalassiosira oceanica} \\ 
      & & NC\_015403.1 & \textit{Fistulifera sp} \\ 
      & & NC\_016731.1 & \textit{Synedra acus} \\ 
      & & NC\_016735.1 & \textit{Fucus vesiculosus} \\ 
      & & NC\_018523.1 & \textit{Saccharina japonica} \\ 
      & & NC\_020014.1 & \textit{Nannochloropsis gadtina} \\ 
      \hline
      % Entering second group
      \parbox[t]{1mm}{\multirow{3}{*}{\rotatebox[origin=c]{90}{F1}}} &
      \multirow{3}{*}{3} & NC\_000925.1 & \textit{Porphyra purpurea} \\ 
       &  & NC\_001840.1 & \textit{Cyanidium caldarium} \\ 
      & & NC\_006137.1 & \textit{Gracilaria tenuistipitata} \\ 
      \hline
      % Entering third group
      \parbox[t]{1mm}{\multirow{17}{*}{\rotatebox[origin=c]{90}{Green Algae}}} &
      \multirow{17}{*}{17} & NC\_000927.1 & \textit{Nephroselmis olivacea} \\ 
      & & NC\_002186.1 & \textit{Mesotigma viride} \\ 
      & & NC\_005353.1 & \textit{Chlamydomonas reinhardtii} \\ 
      & & NC\_008097.1 & \textit{Chara vulgaris} \\ 
      & & NC\_008099.1 & \textit{Oltmannsiellopsis viridis} \\ 
      & & NC\_008114.1 & \textit{Pseudoclonium akinetum} \\ 
      & & NC\_008289.1 & \textit{Ostreococcus tauri} \\ 
      & & NC\_008372.1 & \textit{Stigeoclonium helveticum} \\ 
      & & NC\_008822.1 & \textit{Chlorokybus atmophyticus} \\ 
      & & NC\_011031.1 & \textit{Oedogonium cardiacum} \\ 
      & & NC\_012097.1 & \textit{Pycnococcus provaseolii} \\ 
      & & NC\_012099.1 & \textit{Pyramimonas parkeae} \\ 
      & & NC\_012568.1 & \textit{Micromonas pusilla} \\ 
      & & NC\_014346.1 & \textit{Floydiella terrestris} \\ 
      & & NC\_015645.1 & \textit{Schizomeris leibleinii} \\ 
      & & NC\_016732.1 & \textit{Dunaliella salina} \\ 
      & & NC\_016733.1 & \textit{Pedinomonas minor} \\ % 
      \hline
      % Entering fourth group
      \parbox[t]{1mm}{\multirow{3}{*}{\rotatebox[origin=c]{90}{F2}}} &
      \multirow{3}{*}{3} & NC\_001319.1 & \textit{Marchantia polymorpha} \\ 
       &  & NC\_004543.1 & \textit{Anthoceros formosae} \\ 
      & & NC\_005087.1 & \textit{Physcomitrella patens} \\ %
      \hline
      % Entering fifth group
      \parbox[t]{1mm}{\multirow{2}{*}{\rotatebox[origin=c]{90}{F3}}} &
      \multirow{2}{*}{2} & NC\_014267.1 & \textit{Kryptoperidinium foliaceum} \\ 
      & 
      & NC\_014287.1 & \textit{Durinskia baltica} \\ 
      \hline

      % Entering sixth group
      \parbox[t]{1mm}{\multirow{2}{*}{\rotatebox[origin=c]{90}{F4}}}     &
      \multirow{2}{*}{2} & NC\_001603.2 & \textit{Euglena gracilis} \\ 
      &  & NC\_020018.1 & \textit{Monomorphina aenigmatica} \\ 
       \hline
       % Entering seventh group
       \parbox[t]{1mm}{\multirow{5}{*}{\rotatebox[origin=c]{90}{Ferns}}}     &       \multirow{5}{*}{5}
       & NC\_003386.1 & \textit{Psilotum nudum} \\ 
      & & NC\_008829.1 & \textit{Angiopteris evecta} \\
      &  & NC\_014348.1 & \textit{Pteridium aquilinum} \\ 
      & & NC\_014699.1 & \textit{Equisetum arvense} \\ 
      & & NC\_017006.1 & \textit{Mankyua chejuensis} \\ 
      \hline
    % Entering tenth group
       & & & \\
      \parbox[t]{1mm}{\multirow{1}{*}{\rotatebox[origin=c]{90}{F5}}}
 & 1 & NC\_007288.1 & \textit{Emiliana huxleyi}\\
       & & & \\
 \hline
      % Entering eleventh group
      \parbox[t]{1mm}{\multirow{2}{*}{\rotatebox[origin=c]{90}{F6}}}
       & \multirow{2}{*}{2} & NC\_014675.1 & \textit{Isoetes flaccida} \\
      & & NC\_006861.1 & \textit{Huperzia lucidula} \\
      \hline
    \end{tabular}
  \end{minipage}    
  \begin{minipage}{0.30\textwidth}
   \setlength{\tabcolsep}{4pt}
   \begin{tabular}{|p{0.1cm}|p{0.1cm}|p{1cm}|p{1.85cm}|}
      \hline
      {{F.}}&{{\#}} & {{Acc. No}} & {{Scientific Name}} \\ 
      \hline

      % Entering eighth group
      \parbox[t]{1mm}{\multirow{45}{*}{\rotatebox[origin=c]{90}{Angiosperms}}}  
      & 
      \multirow{45}{*}{45} & NC\_007898.3 & \textit{Solanum lyopersicum} \\
      & & NC\_001568.1 & \textit{Epifagus virginiana} \\ 
      & & NC\_001666.2 & \textit{Zea Mays} \\ 
      & & NC\_005086.1 & \textit{Amborella trichopoda} \\ 
      & & NC\_006050.1 & \textit{Nymphaea alba} \\ 
      & & NC\_006290.1 & \textit{Panax ginseng} \\ 
      & & NC\_007578.1 & \textit{Lactuca sativa} \\ 
      & & NC\_007957.1 & \textit{Vitis vinifera} \\ 
      & & NC\_007977.1 & \textit{Helianthus annuus} \\ 
      & & NC\_008325.1 & \textit{Daucus carota} \\ 
      & & NC\_008336.1 & \textit{Nandina domestica} \\ 
      & & NC\_008359.1 & \textit{Morus indica} \\ 
      & & NC\_008407.1 & \textit{Jasminum nudiflorum} \\ 
      & & NC\_008456.1 & \textit{Drimys granadensis} \\ 
      & & NC\_008457.1 & \textit{Piper cenocladum} \\ 
      & & NC\_009601.1 & \textit{Dioscorea elephantipes} \\ 
      & & NC\_009765.1 & \textit{Cuscuta gronovii} \\ 
      & & NC\_009808.1 & \textit{Ipomea purpurea} \\ 
      & & NC\_010361.1 & \textit{Oenothera biennis} \\ 
      & & NC\_010433.1 & \textit{Manihot esculenta} \\ 
      & & NC\_010442.1 & \textit{Trachelium caeruleum} \\ 
      & & NC\_013707.2 & \textit{Olea europea} \\ 
      & & NC\_013823.1 & \textit{Typha latifolia} \\ 
      & & NC\_014570.1 & \textit{Eucalyptus} \\ 
      & & NC\_014674.1 & \textit{Castanea mollissima} \\ 
      & & NC\_014676.2 & \textit{Theobroma cacao} \\ 
      & & NC\_015830.1 & \textit{Bambusa emeiensis} \\ 
      & & NC\_015899.1 & \textit{Wolffia australiana} \\ 
      & & NC\_016433.2 & \textit{Sesamum indicum} \\ 
      & & NC\_016468.1 & \textit{Boea hygrometrica} \\ 
      & & NC\_016670.1 & \textit{Gossypium darwinii} \\ 
      & & NC\_016727.1 & \textit{Silene vulgaris} \\ 
      & & NC\_016734.1 & \textit{Brassica napus} \\ 
      & & NC\_016736.1 & \textit{Ricinus communis} \\ 
      & & NC\_016753.1 & \textit{Colocasia esculenta} \\ 
      & & NC\_017609.1 & \textit{Phalaenopsis equestris} \\ 
      & & NC\_018357.1 & \textit{Magnolia denudata} \\ 
      & & NC\_019601.1 & \textit{Fragaria chiloensis} \\ 
      & & NC\_008796.1 & \textit{Ranunculus macranthus} \\ 
      & & NC\_013991.2 & \textit{Phoenix dactylifera} \\ 
      & & NC\_016068.1 & \textit{Nicotiana undulata} \\ 
      \hline
      % Entering ninth group
      \parbox[t]{1mm}{\multirow{7}{*}{\rotatebox[origin=c]{90}{Gymnosperms}}}
      & 
      \multirow{7}{*}{7}& NC\_009618.1 & \textit{Cycas taitungensis} \\ 
      & & NC\_011942.1 & \textit{Gnetum parvifolium} \\ 
      & & NC\_016058.1 & \textit{Larix decidua} \\ 
      & & NC\_016063.1 & \textit{Cephalotaxus wilsoniana} \\ 
      & & NC\_016065.1 & \textit{Taiwania cryptomerioides} \\ 
      & & NC\_016069.1 & \textit{Picea morrisonicola} \\ 
      & & NC\_016986.1 & \textit{Gingko biloba} \\ 
      \hline
     \end{tabular}
  \end{minipage}
 
  \scriptsize 
  \noindent where lineages F1, F2, F3, F4, F5, and F6 are 
  \textit{Red Algae},
  \textit{Bryophytes}, 
  \textit{Dinoflagellates},\\
  \textit{Euglena},
  \textit{Haptophytes}, and \textit{Lycophytes} respectively.
  \normalsize
  \end{minipage}
  
  \end{table}
  
\section{Core genomes extraction}
\subsection{Similarity-based approach}\label{sec:simil}
The first method, described below, considers 
a distance-based similarity measure on genes' coding sequences. 
Such an approach requires annotated genomes, like the ones provided by the NCBI website.
\subsubsection{Theoretical presentation}

We start with the following preliminary definition.%~\cite{acgs13:onp}.
\begin{definition}
\label{def1}
Let $A=\{A,T,C,G\}$  be the nucleotides alphabet, and  $A^\ast$ be the
set  of finite  words on  $A$  (\emph{i.e.}, of  DNA sequences).   Let
$d:A^{\ast}\times   A^{\ast}\rightarrow[0,1]$   be   a   function called
similarity measure   on
$A^{\ast}$. Consider a given value $T\in[0,1]$ called a threshold. For
all   $x,y\in  A^{\ast}$,   we   will  say   that  $x\sim_{d,T}y$   if
$d(x,y)\leqslant T$.
\end{definition}

Let be given a \emph{similarity} threshold $T$  and a \emph{similarity measure} $d$.
The method begins by building  an undirected graph 
between all the DNA~sequences $g$ of the set  of genomes as follows:
there is  an edge between $g_{i}$ and $g_{j}$
if  $g_i \sim_{d,T} g_j$ is established.
In other words, vertices are DNA sequences, and two sequences
are connected with an edge is their similarity is larger 
than a predefined threshold. Remark that this graph 
is generally not connected for sufficiently large thresholds.

This graph is further denoted as the ``similarity'' graph.
We thus say that two coding sequences 
$g_i$, $g_j$ are
equivalent with respect to the relation $\mathcal{R}$ if both $g_i$ and 
$g_j$  belong in the same 
connected component (CC) of this similarity graph, \textit{i.e.}, if there is a path between $g_i$ 
and $g_j$ in the graph. To say this another way, if there is a
finite sequence $s_1, ..., s_k$ of vertices (DNA sequences) such that
$g_i$ is similar to $s_1$, which is similar to $s_2$, etc., and $s_k$ is similar
to $g_j$.

It is not hard to see that this relation is an
equivalence relation whereas $\sim$ is not.
Any class for this relation   is  called  a ``gene'' 
in this article,   where   its  representatives
(DNA~sequences)  are the ``alleles''  of this  gene, such abuse of language
being proposed to set our ideas down.  Thus  this first
method   produces   for   each    genome   $G$,   which   is   a   set
$\left\{g_{1}^G,...,g_{m_G}^G\right\}$    of   $m_{G}$    DNA   coding
sequences, the  projection of each sequence according  to $\pi$, where
$\pi$ maps each sequence into its gene (class) according to $\mathcal{R}$. In
other     words,      a     genome     $G$      is     mapped     into
$\left\{\pi(g_{1}^G),...,\pi(g_{m_G}^G)\right\}$.    Note    that    a
projected genome has no duplicated gene since it is a set.

Consequently, the core  genome (resp.,  the pan genome)  of two genomes
$G_{1}$  and $G_{2}$  is defined  as  the intersection  (resp., as  the
union) of their projected  genomes.  We finally consider the intersection
of  all the  projected genomes,  which  is the  set of  all the  genes
$\dot{x}$  such  that   each  genome  has  at  least   one  allele  in
$\dot{x}$. This set will constitute the core genome of the whole species
under consideration. The  pan genome is computed  similarly as the  union of all
the projected  genomes. 

Remark finally that this first method requires the calculation of all similarities between all allele sequences in all species under consideration. So, even in the case of $\approx 100$ organisms 
and with a focus on a specific family or function, this is a computationally heavy operation.
In a future work, the authors' intention is to take benefits from the very large set of 
already computed similarities, to develop heuristic approaches using this basis of 
knowledge, specific to chloroplastic genes,
making it possible to build rapidly  this similarity graph. 

\subsubsection{Case study}

Let us now consider the 99 chloroplastic genomes introduced earlier.
We will use in this case study either the coding sequences 
downloaded from NCBI website or the sequences predicted by DOGMA. 
DOGMA, which stands for \textit{Dual  Organellar GenoMe Annotator}, has
already been evoked in this article. This is a
 tool  developed in 2004 at  University of Texas    for annotating plant
chloroplast and  animal mitochondrial genomes.  This tool  
translates  a  genome  in all  six  reading frames  and then
queries  its own     amino     acid     sequence     database     using
Blast  (blastx~\cite{altschul1990basic})  with  various ad hoc
parameters. The choice of DOGMA is natural, as this annotation tool is
reputed and specific to chloroplasts. 

Each genome is thus constituted
by a list of coding sequences. In this illustration study,
we have evaluated the similarity between two sequences by
using a global alignment. More precisely, the measure $d$ 
introduced above is the similarity score provided after
a Needleman-Wunch global alignment, as obtained by running
the \emph{needle} command from the \emph{emboss} package
released by EMBL~\cite{Rice2000}. Parameters of the \emph{needle}
command are the default ones: 10.0 for gap open penalty and 0.5
for  gap extension.
\begin{figure}[ht]
\centering
\includegraphics[width=.9\linewidth]{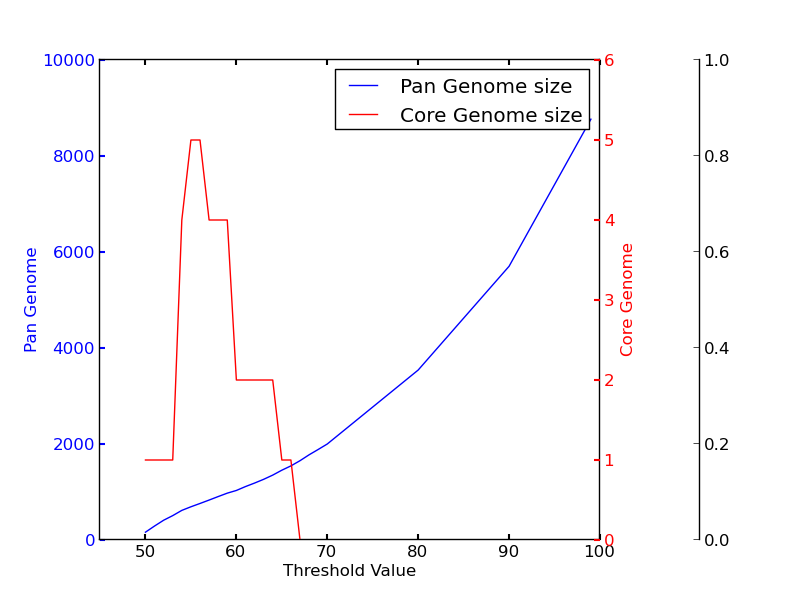}
\caption{Results based on NCBI annotation}
\label{fig:ncbi}
\end{figure}
\begin{figure}[ht]
\centering
\includegraphics[width=.9\linewidth]{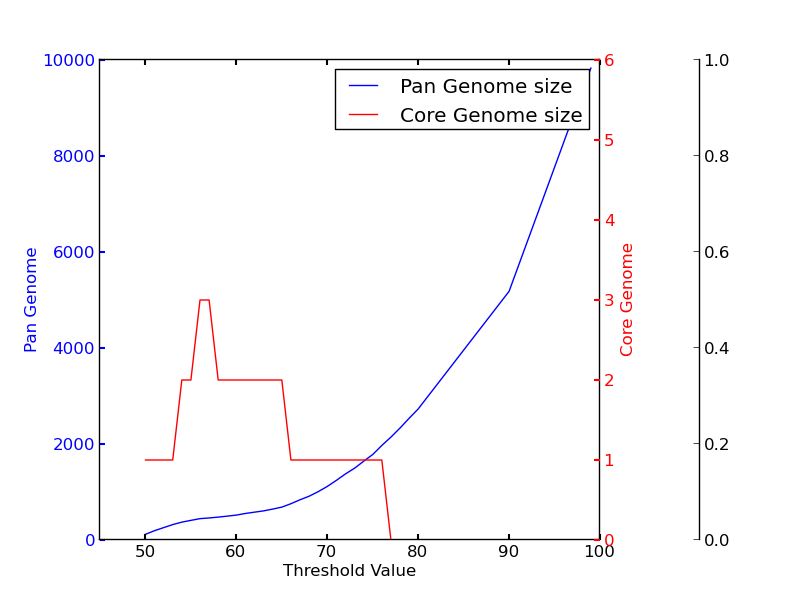}
\caption{Results based on DOGMA annotation}
\label{fig:dogma}
\end{figure}
The number of genes in the core genome and in the pan genome, 
according to this first method using data and measure described above,
have been computed using the supercomputer facilities of the M\'esocentre
de calcul de Franche-Comt\'e. Obtained results are
represented in  
 Figures~\ref{fig:ncbi} and \ref{fig:dogma} with respect to various
threshold values on Needleman-Wunsch similarity scores. 
Remark that when the threshold is large, 
we obtain more connected components, but with small sizes (a large number
of genes, with a few numbers of alleles for each of them). In other words,
when the threshold is large, the 
pan genome is large too.  
No matter the chosen annotation tool, this first approach  suffers from producing
too small core genomes,  for any chosen similarity threshold, compared
to   what  is   usually   expected  by   biologists.

For NCBI, it is certainly due to a wrong determination of start and stop
codons in some annotated genomes, due to a large variety of annotation 
tools used during genomes submission on the NCBI server, some of them
being old or deficient: such truncated 
genes will not produce a large similarity score with their orthologous genes
present in other genomes. The case of DOGMA is more 
difficult to explain as, according to our experiments and to the
state of the art, this gene prediction tool produces normally good
results in average. The best explanation of such an underperformance
is that a few genomes are very specific and far from the remainder ones, in terms
of gene contents, which leads to a small number of genes in the global core
genome. However this first approach cannot help us to determine
which genomes must be removed from our %\cancel{set of} 
data. To do so, we
need to introduce a second approach based on gene names: from the
problematic gene names, we will be able to trace back to the
problematic genomes.

\subsection{Annotation-based approach}\label{sec:annot}
\subsubsection{Using genes names provided by annotation tools}

Instead of using the sequences predicted by annotation tools, we can
try to use the names associated to these sequences, when available.
The basic idea is thus to annotate all the sequences using a given
software, and to consider as a core gene each sequence whose name can
be found in all the genomes.
Two annotation  techniques will be used in the remainder of this article,
namely DOGMA and NCBI. 

It is true that the NCBI annotations are of varying qualities, and sometimes such annotations are totally erroneous. As stated before, it is due to the large variety of annotation tools that can been used during each sequence submission process. However, we also considered it in this article, as this database contains human-curated annotations. To say this another way, DOGMA automatic annotations are good in average, while NCBI contains very good human-based annotations together with possibly bad annotated genomes. 
Let us finally remark that DOGMA also predicts the locations of \textit{ribosomal RNA (rRNA)}, while they are not provided in gene features from NCBI. Thus core genomes constructed on NCBI data will not contain rRNA.

We now investigate core and pan genomes design using each of the two tools separately, which will constitute the second approach detailed in this article. From now on we will consider annotated genomes: either ``genes features'' downloaded from the NCBI, or the result of DOGMA.

\subsubsection{Names processing}

As DOGMA is a deterministic annotation tool, when a given gene is detected twice in two genomes, the same name will be attached to the two coding sequences: DOGMA spells exactly in the same manner the two gene names. So each genome is replaced by a list of gene names, and finding the common core genes between %\cancel{to}
two genomes simply consists in %\cancel{in}
intersecting the two lists of genes. The sole problem we have detected using DOGMA on our chloroplasts is the case of the RPS12 gene: some genomes contain RPS12\_3end or RPS12\_5end in %\cancel{the}
DOGMA result. We have manually considered that all these representatives belong to the same gene, namely to RPS12.

Dealing with NCBI names is more complicated, as various annotation tools have been used together with human annotations, and because there is no spelling rule for gene names. For instance, NAD6 mitochondrial gene is
sometimes written as ND6, while we can find RPOC1, RPOC1A, and RPOC1B in our chloroplasts. So if we simply consider NCBI data without treatment, intersecting two genomes provided as list of gene names often leads to duplication of misspelled genes. Automatic names homogenization is thus required on NCBI annotations, the question being where to draw the line on correcting errors in the spelling of genes? In this second approach, we propose to automate only obvious modifications like putting all names in capital letters and removing useless symbols as ``\_'', ``('', and ``)''.
Remark that such simple renaming process cannot tackle with the situations of NAD6 or RPOC1 evoked above. To go further in automatic corrections requires to use edit distances like Levenshtein, however such use will raise false positives (different genes with close names will be homogenized). The use of edit distances on gene names, together with a DNA sequence validation stage, will be investigated in a second methodology article.

At this stage, we consider now that each genome is mapped to a list of gene names, where names have been homogenized in the NCBI case.

\subsubsection{Core genes extraction}

To extract core genes, we  iteratively collect the maximum  number of common  genes among genomes,  therefore   during  this  stage an \textit{Intersection Core Matrix} (ICM) is built.   ICM is  a two dimensional symmetric matrix where each row and each column corresponds to   one   genome.   Hence,   an   element   of   the  matrix   stores the  \textit{Intersection Score}  (IS):  the cardinality  of the  core genes obtained by intersecting the two genomes.
Mathematically speaking, if we have $n$ genomes in local database, the ICM is an $n \times n$ matrix whose elements satisfy: 
\begin{equation}
score_{ij}=\vert g_i \cap g_j\vert
\label{Eq1}
\end{equation}

\noindent where $1 \leq i \leq n$, $1 \leq j \leq n$, and $g_i, g_j$ are genomes. The  generation of a new  core genome depends  obviously on the value  of the  intersection scores  $score_{ij}$. More  precisely, the idea is  to consider a  pair of genomes  such that their score  is the largest element in the ICM. These two genomes are then removed from the matrix and the  resulting new  core genome is  added for the  next iteration.
The ICM is then updated to take into account the new core genome: new IS values are computed for it. This process is repeated until no new core genome can be obtained.

We  can observe  that  the ICM  is relatively  large  due to  the amount  of species. As a consequence, the  computation of the intersection scores is both  time and  memory consuming.  However,  since ICM  is obviously a  symmetric matrix we can reduce the  computation overhead by considering only its upper triangular part.  The  time complexity  for this  process is: $O(\frac{n.(n-1)}{2})$.  Algorithm~\ref{Alg1:ICM} illustrates the construction  of the ICM matrix and  the extraction of the  core  genomes, where  \textit{GenomeList}  represents the  database
storing all genomes  data. At each iteration, this algorithm computes the maximum core genome with its two  parents (genomes).

\begin{algorithm}[ht]
\caption{Extract Maximum Intersection Score}
\label{Alg1:ICM}
\begin{algorithmic} 
\REQUIRE $L \leftarrow \mbox{genomes sets}$
\ENSURE $B1 \leftarrow \mbox{Max Core set}$ 
\FOR{$i \leftarrow 1:len(L)-1$}
        \STATE $score \leftarrow 0$
	\STATE $core1 \leftarrow set(GenomeList[L[i]])$
	\STATE $g1 \leftarrow L[i]$
	\FOR{$j \leftarrow i+1:len(L)$}
	        \STATE $core2 \leftarrow set(GenomeList[L[j]])$
		\STATE $core \leftarrow core1 \cap core2$
		\IF{$len(core) > score$}
                  \STATE $score \leftarrow len(core)$
		  \STATE $g2 \leftarrow L[j]$
                \ENDIF
	\ENDFOR
	\STATE $B1[score] \leftarrow (g1,g2)$
\ENDFOR
\RETURN $max(B1)$
\end{algorithmic}
\end{algorithm}
\subsubsection{Features visualization}
The last stage of the proposed pipeline is naturally to take advantage
of the produced core and pan genomes for biological studies. As
this key stage is not directly related to the methodology for core
and pan genomes discovery, we will only outline a few tasks that
can be operated on the produced data.
\begin{figure}[H]
\begin{frame}
\centering
\includegraphics[width=1.05\linewidth]{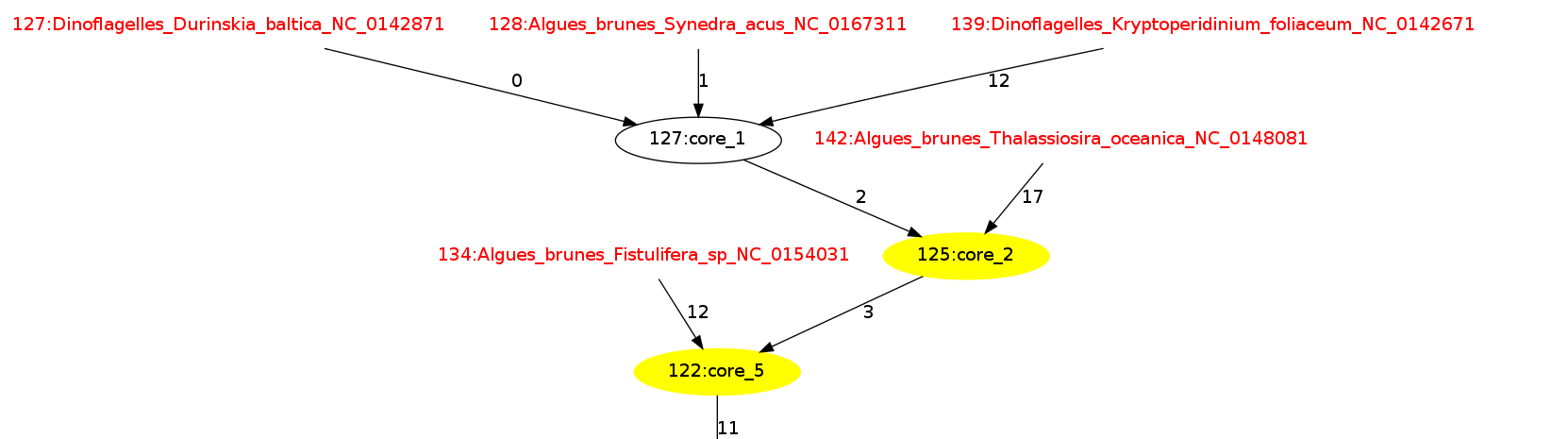}
\caption{Part of a core genomes evolutionary tree (NCBI gene names)}
\label{coreTree}
\end{frame}
\end{figure}

Obtained results may be visualized by building a core genomes evolutionary tree, simply called \textit{core tree}. Each node in this tree represents a chloroplast genome or a predicted core,
as depicted in Figure~\ref{coreTree}. In this figure, nodes labels are of the form \textit{(Genes number:Family name\_Scientific name\_Accession number)}, while an edge is labeled
with the number of gene loss when compared to its parents (a leaf genome or an intermediate core genome). Such numbers can answer questions like: how
many genes are different between two species? Which functionality has been lost between an ancestor and its children? For complete core trees based either on NCBI names or on DOGMA ones, see http://members.femto-st.fr/christophe-guyeux/en/chloroplasts.

A second application of such data is obviously to build accurate phylogenetic trees, using tools like
PHYML\cite{guindon2005phyml} or 
RAxML{\cite{stamatakis2008raxml}. 
Consider a set of species, the least  common core genome in a core tree 
contains all shared common genes among these species. To constitute a 
phylogenetic tree, core genes will  be
multi-aligned to serve as an input to  the phylogenetic tools mentioned above.
An example of such a phylogenetic tree for core 58 is provided in Figure~\ref{phylo1:58}. %(NCBI cores tree, see \url{http://members.femto-st.fr/christophe-guyeux/en/chloroplasts}). 
Remark that, in order to constitute the phylogenetic tree, a relevant outgroup is needed from \emph{Cyanobacteria}. The process simply starts by blasting each gene in the core with outgroup genes, and then selects the relevant one. 

\begin{figure}[H]
\centering
\includegraphics[width=1.1\linewidth]{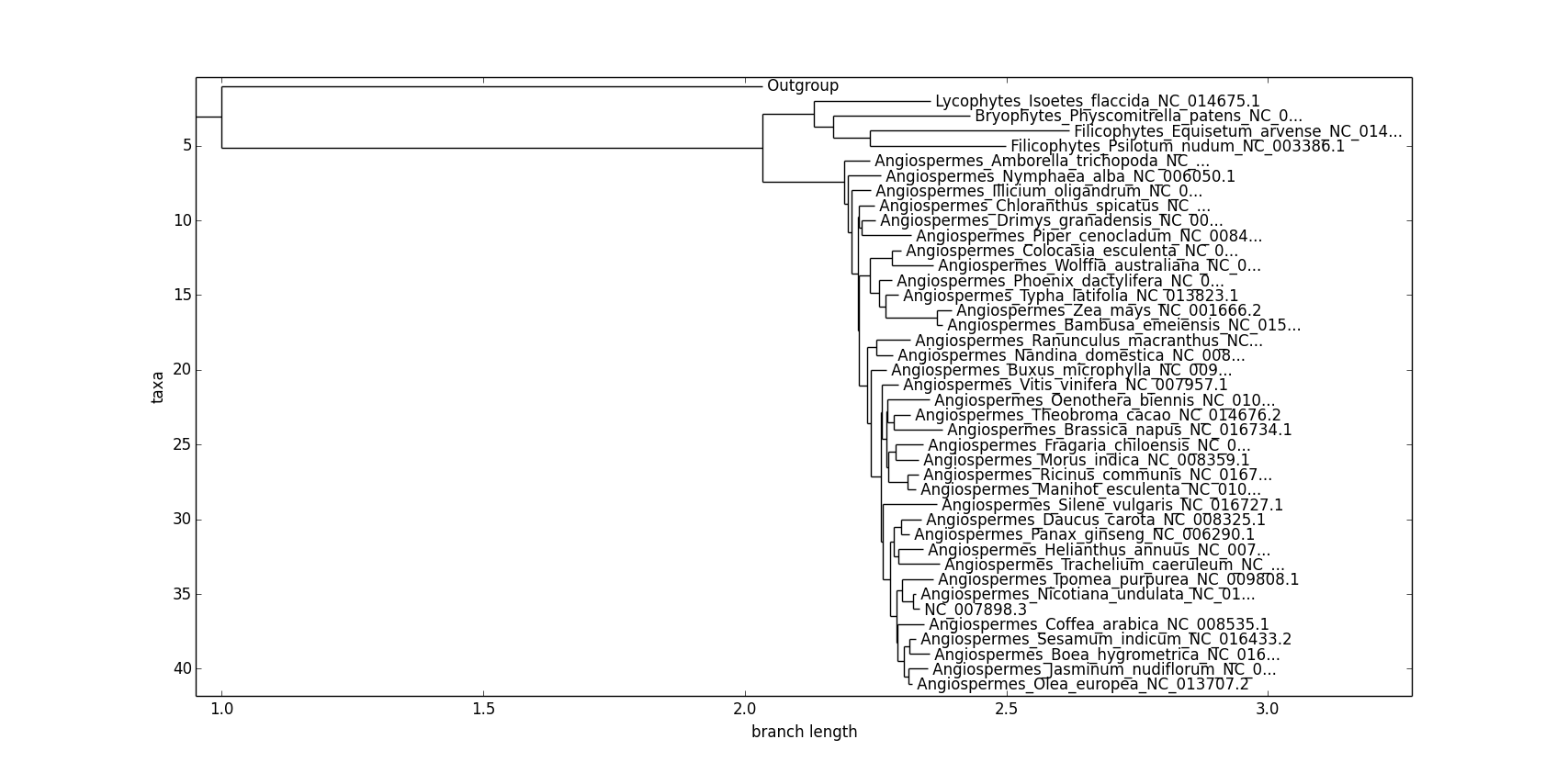}
\caption{Part of a phylogenetic tree for core 58 (NCBI gene names)}
\label{phylo1:58}
\end{figure}
\section{Discussion}
\subsection{Biological evaluation}
\label{sec:discuss}
It is well known that the first plants' endosymbiosis ended in a great diversification of lineages comprising \textit{Red Algae}, \textit{Green Algae}, and \textit{Land Plants} (terrestrial). Several second endosymbioses occurred then: two involving a \textit{Red Algae} 
and other heterotrophic eucaryotes and giving birth to both \textit{Brown Algae} and \textit{Dinoflagellates} lineages; another involving a \textit{Green Algae} and a heterotrophic eucaryote and giving birth to \textit{Euglens}~\cite{mcfadden2001primary}.

The interesting point with the produced core trees (especially the one 
obtained with DOGMA, see~\url{http://members.femto-st.fr/christophe-guyeux/en/chloroplasts}) is that organisms resulting from the first endosymbiosis are distributed in each of the lineages found in the chloroplast genome structure evolution. More precisely, all \textit{Red Algae} chloroplasts are grouped together in one lineage, while \textit{Green Algae} and \textit{Land Plants} chloroplasts are all in a second lineage. 
Furthermore organisms resulting from the secondary endosymbioses are well localized in the tree: both the chloroplasts of \textit{Brown Algae} and \textit{Dinoflagellates} representatives are found exclusively in the lineage also comprising the \textit{Red Algae} chloroplasts from which they evolved, while the \textit{Euglens} chloroplasts are related to the \textit{Green Algae} chloroplasts from which they evolved. This makes sense in terms of biology, history of lineages, and theories of chloroplasts origins (and so photosynthetic ability) in different Eucaryotic lineages~\cite{mcfadden2001primary}.

Interestingly, the sole organisms under consideration that possess a chloroplast (and so a chloroplastic genome) but that have lost the photosynthetic ability (being parasitic plants) are found at the basis of the tree, and not together with their phylogenetically related species. This means that functional chloroplast genes are evolutionary constrained when used in photosynthetic process, but loose rapidly their efficiency when not used, as recently observed for a species of Angiosperms~\cite{li2013complete}. 
These species are \textit{Cuscuta gronovii}, an Angiosperm (flowering plant) at the base of the DOGMA Angiosperm-Conifers branch, and 
\textit{Epifagus virginiana}, also an Angiosperm, at the complete basis of this tree.

Another interesting result is that \textit{Land Plants} that represent a single sub-lineage originating from the large and diverse lineage of \textit{Green Algae} in Eucaryotes history are present in two different 
branches of the DOGMA tree, both associated with \textit{Green Algae}: one branch comprising the basal grade of \textit{Land Plants} (mosses and ferns) and the second one containing the most internal lineages of \textit{Land Plants} (Conifers and flowering plants). But independently of their split in two distinct branches of the DOGMA tree, the \textit{Land Plants} always show a larger number of functional genes in 
their chloroplasts than the \textit{Green Algae} from which they emerged, probably meaning that the terrestrial way of life necessitates more functional genes for an optimal photosynthesis than the marine one. However, a more detailed analysis of selected genes is necessary to better understand the reasons why such a distribution has been obtained.
Remark finally that all these biologically interesting results are apparent only in the core tree based on DOGMA, while they are not so obvious in the NCBI one. 

\section{Conclusion} \label{sec:concl}
In this research work, we studied two methodologies for extracting core genes from a large set of chloroplasts genomes, and we developed 
Python programs to evaluate them in practice. 

We firstly considered to extract core genomes by the way of comparisons 
(global alignment) of DNA sequences downloaded from NCBI database. 
However this method failed to produce biologically 
relevant core genomes, no matter the chosen similarity threshold, probably
due to annotation errors. We then considered to use the DOGMA annotation tool 
to enhance the genes prediction process. The second method consisted in extracting 
gene names either from NCBI gene features or from DOGMA results. A first
``intersection core matrix (ICM)'' were built, in which each coefficient 
stored the intersection cardinality of the two genomes placed at the extremities
of its row and column. New ICMs are 
then successively constructed by selecting the maximum intersection score (IS)  in this matrix, 
removing each time the two genomes having this score, and adding the corresponding 
core genome in a new ICM construction. 

Core trees have finally been generated for each method, to investigate 
the distribution of chloroplasts and core genomes. The tree from second 
method based on DOGMA has revealed the best distribution of
 chloroplasts regarding their evolutionary history. In particular, it appears to
us that each endosymbiosis event is well branched in the DOGMA core tree.     

In future work, we intend to deepen the methodology evaluation by considering
new gene prediction tools and various similarity measures on both
gene names and sequences. Additionally, we will investigate new clustering
methods on the first approach, to improve the results quality in this promising way to 
obtain core genes. Finally, the results produced with DOGMA will be 
further investigated, biologically speaking: the genes content of each core
will be studied while phylogenetic relations between all these species
will be questioned.
\bigskip

\textit{Computations have been performed on the supercomputer facilities of the M\'esocentre
de calcul de Franche-Comt\'e.}

\bibliographystyle{splncs}

\begin{IEEEbiography}[{\includegraphics[width=1in,height=1.25in,clip,keepaspectratio]{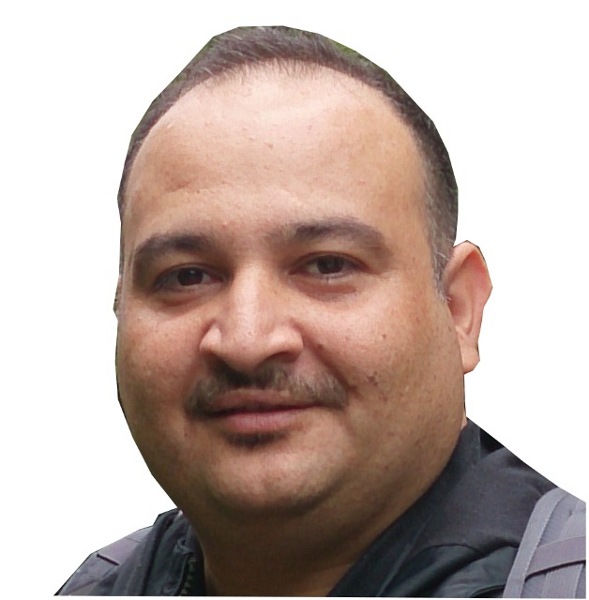}}]{Bassam ALKINDY}
was born in january 1978 in Baghdad, Iraq. In 2005, he defended his M.Sc. thesis in Robotics, from the University of Yarmouk, faculty of information technology and computer science, department of Computer Science, Jordan.

Since 2006, he became a full time lecturer in the University of Mustansiriyah, Baghdad-Iraq. Since 2012, he became a Ph.D. student in the University of Franche-Comté, Besançon-France, in Femto-ST/ DISC - Department of computer science. He is interesting in the domains of Artificial Intelligent, bioinformatics, and Machine learning. 
\end{IEEEbiography}
\begin{IEEEbiography}[{\includegraphics[width=1in,height=1.25in,clip,keepaspectratio]{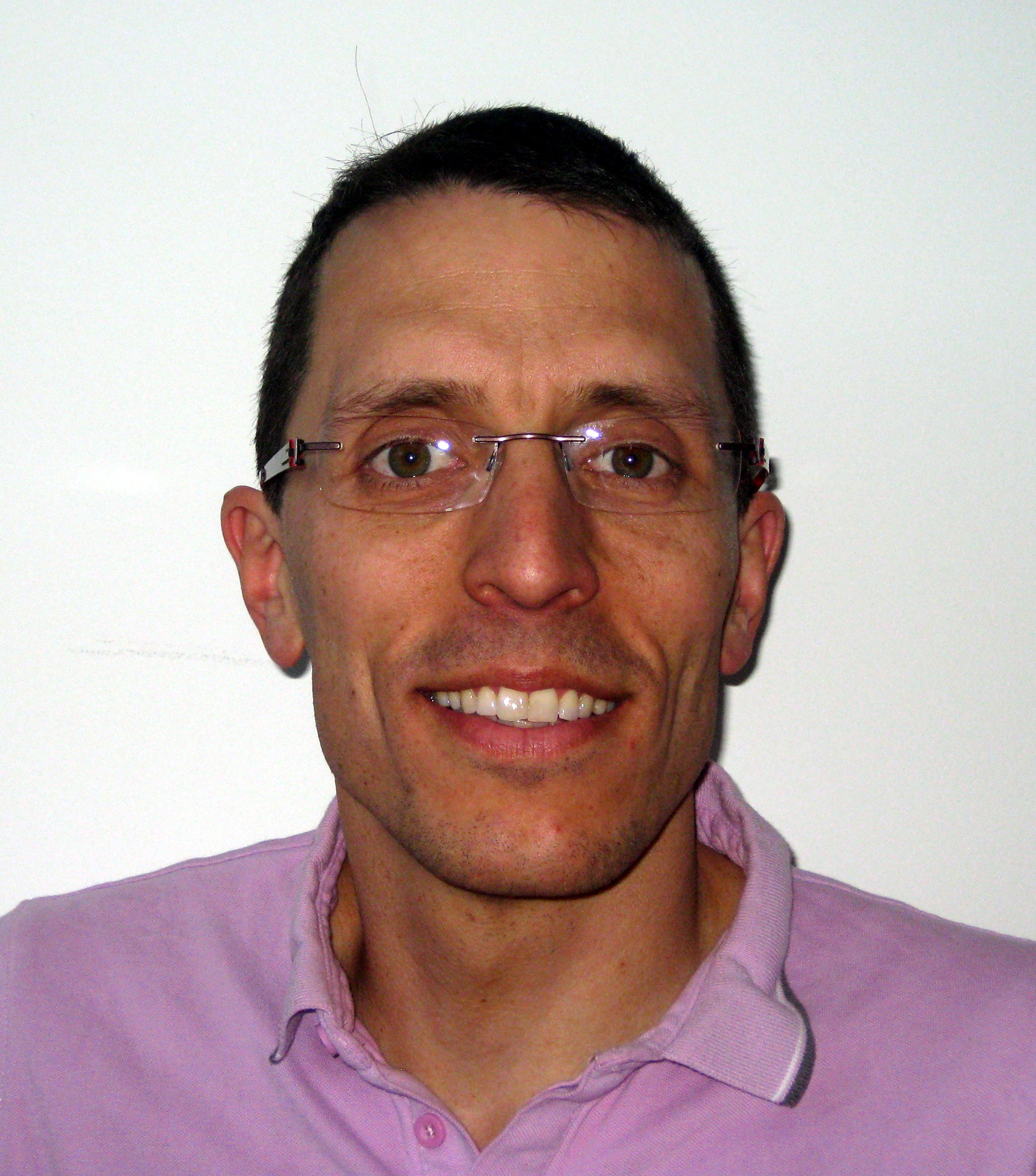}}]{Jean-Fran\c{c}ois Couchot}
is an Assistant Professor in the Department of Computer Science (DISC) of the FEMTO-ST Institute (UMR 6174 CNRS) at the University of Franche-Comté. He received a Ph.D. in Computer Science in  2006 in the FEMTO-ST Institute and applied for a postdoctoral position at INRIA Saclay Île de France in 2006. His research focuses on discrete dynamic systems (with applications in data hiding, pseudorandom number generators, hash function) and on bioinformatics, especially in genes evolution prediction.
He has written more than 20 scientific articles in these areas.

\end{IEEEbiography}
\begin{IEEEbiography}[{\includegraphics[width=1in,height=1.25in,clip,keepaspectratio]{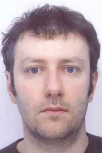}}]{Christophe Guyeux}
has taught mathematics and computer science in the Belfort-Montbéliard university Institute of Technologies (IUT-BM) this last decade.

He has defended a computer science thesis dealing with security, chaos, and dynamical systems in 2010 under Jacques Bahi’s leadership, and is now an associated professor in the computer science department of complex system (DISC), FEMTO-ST Institute, University of Franche-Comté. Since 2010, he has published 2 books, 17 articles in international journals, and 27 articles dealing with security, chaos, or bioinformatics.
\end{IEEEbiography}
\begin{IEEEbiography}[{\includegraphics[width=1in,height=1.25in,clip,keepaspectratio]{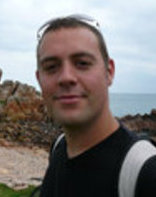}}]{Arnaud Mouly}
was born in France, in October 1978. He completed his Ph.D studies at the National Herbarium of the Nation Museum of Natural History. After the Ph.D degree he occupied a postdoc position in systematic botany at the Bergius Foundation of Royal Swedish Academy of Sciences.
He is currently assistant professor in plant systematics and ecology at the University of Franche-Comté, in Besançon, France. His research interest includes botany, diversity dynamics in insular systems, and usages of phylogenetic trees to answer biological/ecological questions. He is also the director of the Botanical garden of Besançon, France.
\end{IEEEbiography}
\begin{IEEEbiography}[{\includegraphics[width=1in,height=1.25in,clip,keepaspectratio]{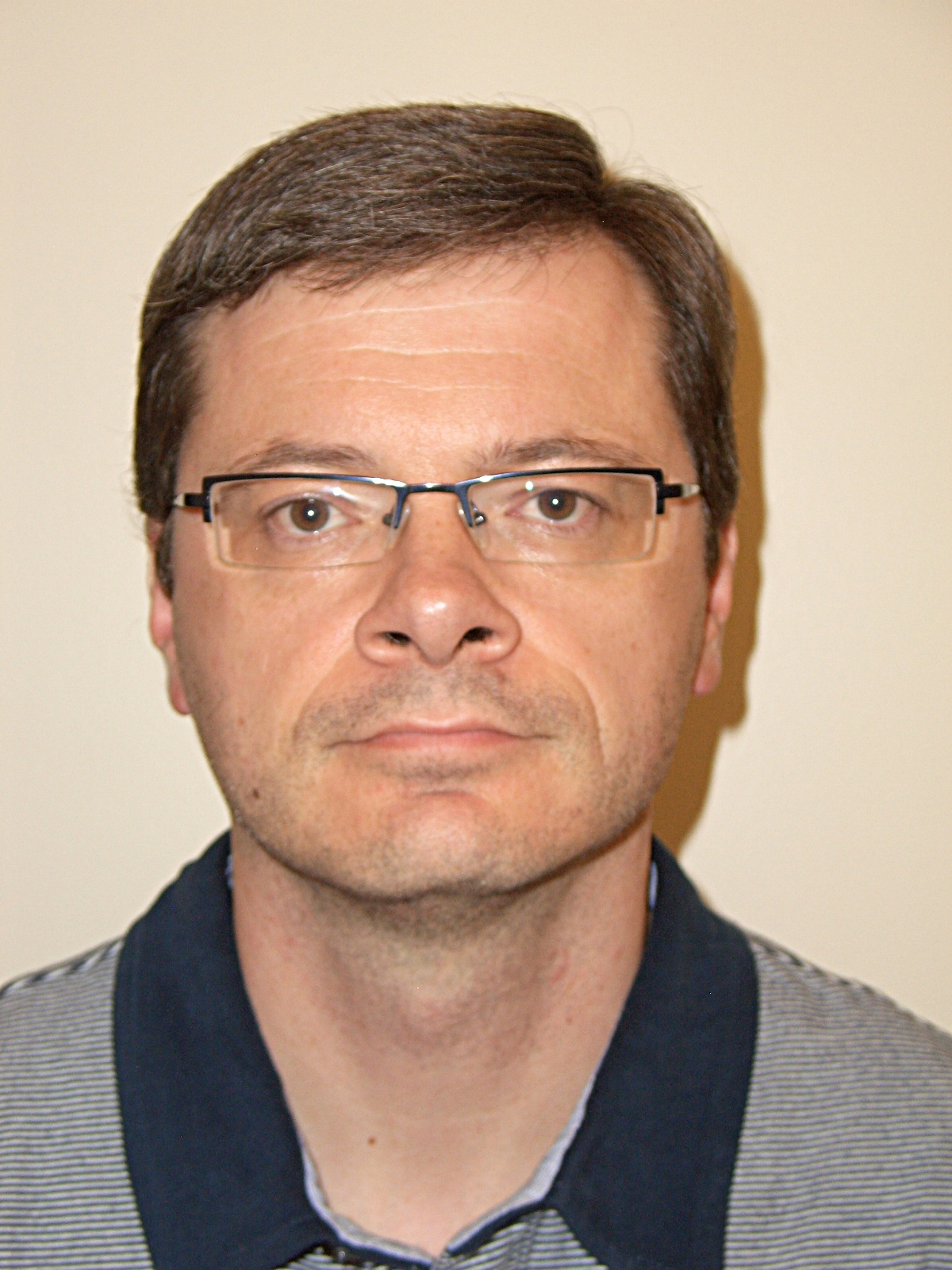}}]{Michel Salomon}
is an Assistant Professor in the Department of Computer Science (DISC) of the FEMTO-ST Institute (UMR 6174 CNRS-UFC-ENSMM-UTBM) at the University of Franche-Comté (UFC). He received a Ph.D. in Computer Science in December of 2001 from the University of Strasbourg. His research focuses on dynamic systems, in particular machine learning approaches, with applications in various areas: wireless networks, radiotherapy (prediction of respiratory motion and the dose deposited), bioinformatics (protein folding and so on.), or active airflow control.
\end{IEEEbiography}
\begin{IEEEbiography}[{\includegraphics[width=1in,height=1.25in,clip,keepaspectratio]{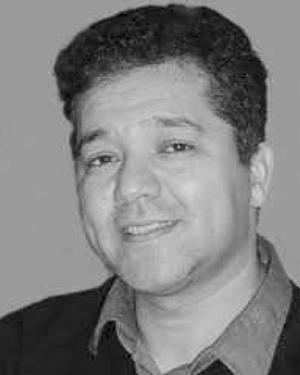}}]{Jacques M. Bahi}
was born in July, 25th 1961. He received the M.Sc. and Ph.D. degrees in applied mathematics from the University of Franche-Comté, France, in 1991.

From 1992 to 1999, he was an Associate Professor of applied mathematics at the Mathematical Laboratory of Besançon. His research interests were focused on parallel synchronous and asynchronous algorithms for differential algebraic equations and singular systems. Since September 1999, he has been a full professor of computer science at the University of Franche-Comté. He published about 150 articles in peer reviewed journal and international conferences and 2 scientific books.

Dr. Bahi is the head of the Distributed Numerical Algorithms team of the Computer Science Laboratory of Besançon, he supervised 21 Ph.D. students. He is a member of the editorial board of 2 international journals and is regularly a member of the scientific committees of many international conferences. Currently, he is interested in: 1) high performance computing, 2) distributed numerical algorithms for ad-hoc and sensor networks and 3) dynamical systems with application to data hiding and privacy.
\end{IEEEbiography}
\end{document}